\documentclass[a4paper,11pt]{article}
\usepackage{pos}
\usepackage{graphicx}   

\title{Polarized muons and the origin of biological homochirality}
 \ShortTitle{Polarized muons and  biological homochirality}

\author*[a,b,c,d]{No\'emie Globus}
\author[e]{Roger D. Blandford}
\author[f]{Anatoli Fedynitch}

\affiliation[a]{Department of Astronomy and Astrophysics, University of California, Santa Cruz, CA 95064, USA. E-mail: noglobus@ucsc.edu}
\affiliation[b]{Center for Computational Astrophysics, Flatiron Institute, Simons Foundation, New-York, NY 10003, USA}
\affiliation[c]{ELI Beamlines, Institute of Physics, Czech Academy of Sciences, 25241 Dolní Brěžany, Czech Republic}
\affiliation[d]{Astrophysical Big Bang Laboratory (ABBL), Cluster for Pioneering Research, RIKEN, 21 Hirosawa, Wakō, Saitama 3510198, Japan}
\affiliation[e]{Kavli Institute for Particle Astrophysics \& Cosmology, Stanford University, Stanford, CA 94305, USA. E-mail: rdb3@stanford.edu}
\affiliation[f]{Institute for Cosmic Ray Research, the University of Tokyo, 5-1-5 Kashiwa-no-ha, Kashiwa, Chiba 277-8582, Japan. E-mail: afedyni@icrr.u-tokyo.ac.jp}

\abstract{While biologists have not yet reached a consensus on the definition of life,   homochirality - the specific molecular handedness of biomolecules -  is a phenomenon only produced by life. The unraveling of its origin requires interdisciplinary research, by exploring  fundamental physics, chemistry, astrophysics and biology. Here, we consider the origin of biological homochirality in the context of astrophysics and particle physics. The weak force, one of the fundamental forces operating in nature, is parity-violating.  On Earth, at ground level, most of our cosmic radiation dose comes from polarized muons formed in a decay involving the weak force. We discuss how the  magnetic polarization is transmitted in cosmic showers within several different environments which are prime targets in the search for the origin of life. We consider how this polarization could have induced a biological preference for one type of chirality over the other, and  discuss the implications for the search of life in other worlds.}

\FullConference{37$^{\rm{th}}$ International Cosmic Ray Conference (ICRC 2021)\\
		July 12th -- 23rd, 2021\\
		Online -- Berlin, Germany}

\begin{document}
\maketitle
\section{Cosmic muons, an important but overlooked factor in astrobiology}
The search for life beyond Earth requires an understanding of the environments that support it. One key environmental factor is the background ionizing radiation that can be of terrestrial, biological, or cosmic origin. Today, at sea level,, the dose of ionizing radiation  to organisms is dominated by the naturally-occurring 
isotopes of Uranium, %($^{235}$U half life $703.8 \times 10^6$ years, $^{238}$U half life $4.47 \times 10^9$ years), 
Thorium  and Potassium  %($^{40}$K half life $1.25 \times 10^9$ years),
 and their decay products, either by inhalation, ingestion  or simply due to external exposure. The dose  has changed over time due to the evolution of the continental crust, and changes in the ratios and abundance of these radioactive elements. Not accounting for human-made radiation, the exposure due to natural radioactivity  constitute $\sim$90\% of the total background radiation dose. The remaining $\sim$10\%  comes from space radiation, in the form of ultraviolet light or cosmic rays. 
 
 We are immersed in a sea of cosmic rays coming from the Sun (energies $\lesssim$ GeV), from Galactic supernovae  occurring in the spiral arms  (energies $\lesssim$ 100 TeV), and from rare "Pevatrons" that  mostly populate the center of our Galaxy and  can accelerate cosmic ray to a few PV/c rigidities. Cosmic rays with higher rigidities come from unknown sources whose nature is  debated at this conference. The highest energy cosmic rays, with rigidities $\gtrsim 1$ EV/c, are barely deflected by the large scale magnetic fields of the Milky Way halo, and are likely of  extragalactic origin; but they are rare (with a flux of less than one particle per square kilometer per year) and, unless some rare catastrophic event were to occur in our local neighborhood,  the cosmic-ray flux over geological times is largely dominated by the Galactic cosmic rays. 
 
\begin{figure}
\centering
       \includegraphics[scale=0.48]{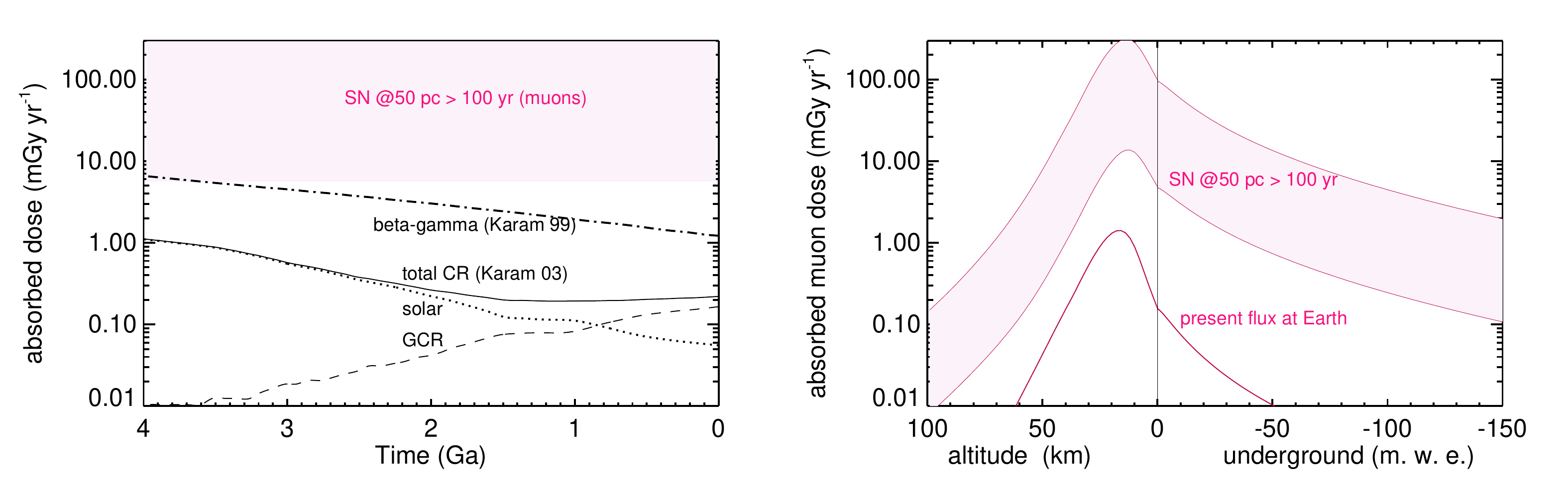}
\caption{ Left panel: Radiation dose rate from geological and biological radionuclides through time, taken from \cite{karam99}, and evolution of the cosmic radiation dose rate, taken from \cite{karam03}, compared to the muon radiation dose (pink shaded area) enhancement due to a supernova at 50 pc. Right panel: Absorbed dose at Earth for the presently observed cosmic ray flux according to the GSF model \cite{Dembinski17}  and for a nearby supernova, 100
years after explosion at 50pc \cite{melott17}. The upper edge represents case A from \cite{melott17} and the lower case B, respectively. Shown is the maximal fluxes expected 100 years after explosion that drop significantly within a few kyr. Right panel figure adapted from \cite{GFB21}.}
\label{fig:doses}
\end{figure}

 The average cosmic radiation flux over long intervals of time ($>$1~Myr) appears to be mostly stable \cite{lavielle99}. The left panel of  Figure~\ref{fig:doses} shows the evolution of the radiation doses due to cosmic radiation compared to natural radioactivity over geological times \cite{karam99,karam03}. When the very first organisms began leaving chemical signs of their existence, $\sim$3.8 billion years ago, the annual beta and gamma radiation doses  was about 7 mSv\footnote{For X rays, gamma rays, muons and beta radiation, the conversion factor between absorbed dose in mGy and equivalent dose in mSv is one,  1 mSv = 1 mGy. So one can use equivently Sv or Gy for these types of radiation.}; It is less than 2 mSv today \cite{karam99}.  For cosmic radiation contributing to the world population-weighted effective dose, recent study suggests 0.32 mSv with a range between 0.23 and 0.70 mSv (and subject to the solar modulation condition within about 15\%) \cite{Sato16}. %, which is slightly smaller than 0.38 mSv with a range between 0.3 and 2 mSv suggested by UNSCEAR 2000 report.}
 The level of cosmic radiation varies with the altitude and latitude because of atmospheric and geomagnetic effects. The long term average can punctuated by increases in cosmic ray solar activity \citep{Winckler58}, by occasional nearby supernovae \cite{Ellis95,thomas16}, by occasional (and rarer) gamma-ray bursts  from core-collapsing stars \cite{globus15} or binary neutron star mergers \cite{gottlieb21}, by the periodic increase during spiral arm passages \cite{shaviv02}, or by the increase in the star formation activity \cite{Svensmark98}. For instance, when a supernova occurs in our local neighborhood, a period of few kilo years following the event will be characterized by a higher level of cosmic radiation, a factor of a hundred to a thousand above the long term average \cite{melott17},  thereby enabling the probability of affecting the atmospheric properties and stressing life.  According to \cite{Wallner16}, a supernova explodes in our local galactic neighbourhood (within a distance of about 100 pc)  every 2 - 4 Myr.

 When cosmic rays interact with planetary atmospheres or the surfaces of  moons, asteroids or comets, they initiate showers of secondary particles: hadrons, electrons, photons, and muons.  On Earth, at ground level, $\sim$85\% of the cosmic radiation dose comes from muons \cite{GB20}. Muons will affect everything except organisms below about 1 kilometer of water (or hundreds of meters of rock); at  one kilometer below the ocean' surface, the muon flux is attenuated by a factor thousand~\cite{Bugaev98}. The radiation dose due to muons below and above ground is shown in the right panel of Figure~\ref{fig:doses}. By comparison, UVB (ultraviolet in the range 290 nm to 320 nm) effects stop at about 10 meters depth of water.  Although ozone depletion effect by UVB  can be important, it would not affect much marine or underground life, whereas a nearby supernova event enhances the  muon radiation dose by a few orders of magnitude \cite{GFB21}, far beyond the dose from natural radioactivity, as shown  by the pink shaded areas in Figure~\ref{fig:doses}. {\bf Therefore, local $(\lesssim50~{\rm pc})$ supernovae would have a dominant impact through their muons \cite{thomas16,melott17}.} %However, the potential role of muons in astrochemistry and astrobiology has been so far overlooked. 
 It is therefore of prime importance to understand the effects of muon radiation on living organisms.  
 
 Living cells exposed to radiation can: (1) repair themselves, leaving no damage; (2) die and be replaced, much like millions of our body cells do every day; or (3) incorrectly repair themselves, resulting in a biophysical change. High radiation doses (greater than 500 mSv) tend to kill cells, while low doses spread out over a long period would not cause an immediate problem because life has time to evolve and adapt. As can be seen on the right panel of Figure~\ref{fig:doses}, the elevated muon dose level from a nearby supernova would not exceed $\sim$500 mGy. While the rare and cataclysmic events such as gamma-ray bursts have been invoked as a limiting factor for life \cite{Ellis95,dar98}, some species on Earth are found to be extremely resistant to radiation. Tardigrades can survive to radiation doses of  5,000 Gy \cite{Horikawa06}. It is possible that the their ability to resist  radiation dose rates far larger than the present exposure  may have been established during earlier periods of high radiation. 
 The oldest remains of modern tardigrades are around 90 million years old. They may have stopped evolving  due to their high resistance to radiation, inhibiting mutation. Another example is the eukaryotic algae found in the spent fuel pool of a research reactor in France,  resisting a dose of ~20,000 Sv - more than 2,000 times the lethal dose to human beings \cite{Farhi08}. 

So, if life can adapt to such high radiation doses, why haven't we found aliens yet? The question is not so much to find environments where life can survive, the question is to find environments where life can \textit{originate}. This is truly the limiting factor, because life at its beginning was not as much armed against ionizing radiation as it is today. This is different from the concept of "habitable zones" (essentially those in which liquid water can exist \cite[][]{Kasting}), so we call the environments  "founts"~\cite{GB20}. To understand where life can form, we can study what are  the fundamental features of life as we know it and search for the associated signatures in our solar system, or on exoplanets.  {\bf The perfect molecular mirror asymmetry exhibited by the living organisms, a property known as "homochirality", is such a biosignature, because is a phenomenon only produced by life. }

\section{Molecular homochirality, an unambiguous signature of life}

Chirality is the geometric property of an object that cannot be superimposed on its mirror image~\citep{kelvin1894}. In chemistry,  mirror images of the same molecule are called enantiomers\footnote{from the Greek $\epsilon\chi\theta\rho$\'o$\varsigma$, "enemy" or "opposite".}.  In 1848, Louis Pasteur recognised that biological molecules make one choice of the two enantiomers \citep{pasteur1848} which allows  proteins and nucleic acids to adopt stable helical structures that are essential to their function. From the chemistry point of view, two mirror-image molecules may possess identical information. However, for living organisms, they are not equivalent since only one of the two enantiomers is used for specific biological functions. The ribonucleic and deoxyribonucleic acids (RNA and DNA), responsible for the replication and storage of genetic information, are made up of linear sequences of  building blocks, called nucleotides, with the same handedness whose arrangement contains the genetic information needed to sustain life \citep{schrodinger1944,shannon1948, watson1953, shinitzky2007}.  

\begin{figure}
\centering
       \includegraphics[scale=0.35]{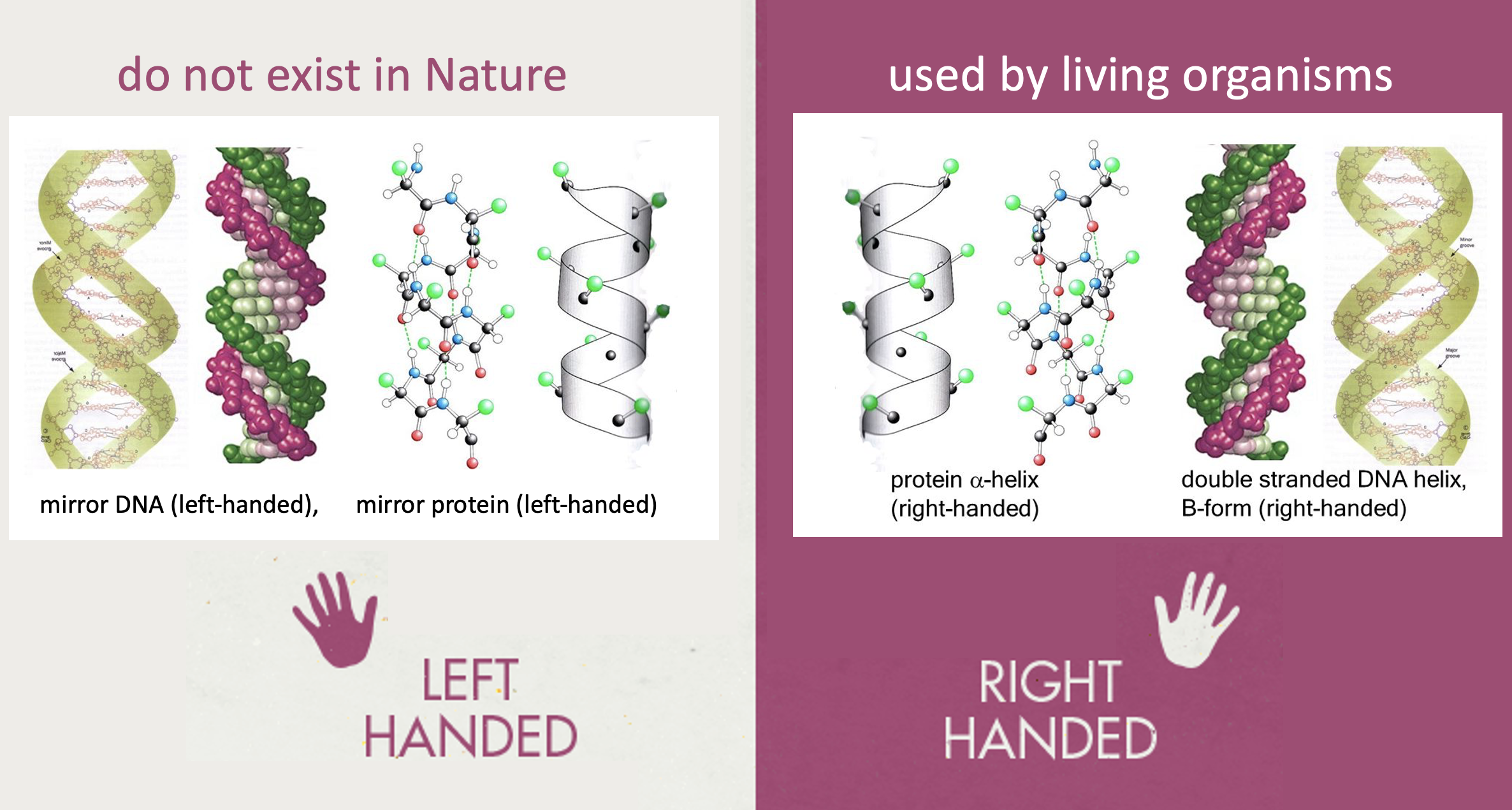}
\caption{The figure shows the 3D structure of the DNA molecule and protein, and their mirror images.   The direction of the helical conformation of the nucleic acids derives from the underlying chemical chirality of the sugar backbone. The nucleic acids  contain only  right-handed sugars (D-ribose in RNA, D-deoxyribose in DNA), shown in the right-hand side of the figure. They naturally assume a right-handed helical conformation.  
In the mirror world (left-hand side in the figure), the nucleic acids would contain only left-handed sugars (L-ribose or L-deoxyribose) and would assume a predominantly  left-handed helical conformation. Note that "left-handed" and "right-handed" are only human conventions.}
\label{fig:chirality}
\end{figure}

The chirality of the sugars and amino acids confers helical structure on nucleic acids and proteins.  As RNA and DNA are made of right-handed sugars ("right-handed" is a human convention), the more stable conformation is a right-handed helix (see Figure \ref{fig:chirality}). The homochirality of the sugars has important consequences for the stability of the helix, and hence, on the fidelity or error control of the genetic code. All the twenty encoded amino acids are left-handed (again by human convention) and this leads to a right-handed structure in proteins, called $\alpha$-helix.  The way a protein folds into space determines its biological function.  Homochirality is thus an unambiguous biosignature and its presence on an extraterrestrial body will be a powerful indicator of life \cite{avnir20,Patty21}. 

 Direct evidence of chiral molecular asymmetry in amino acids   
 has been  found in carbonaceous chondrites \citep{cronin1997},
 the most primitives and least altered meteorites.  The enantiomeric excesses  \cite[up to 15\%, e.g.][]{burton08} show a preference for left-handed amino acids over right-handed ones.  Asymmetries in right-handed sugars over left-handed sugars have also been reported \citep{cooper2016}.  Note that chondrites contain a wide range of extraterrestrial nucleobases \citep{martins2008}, suggesting that meteorites could have delivered  chiral molecules, and maybe even life, to Earth. In 2016, propylene oxide was discovered in interstellar space \citep{mcguire2016}. This was the first detection of a chiral molecule, though there was no evidence that the source was homochiral. Because meteorites samples are subject to Earth contamination, ongoing and future space missions are devoted to measure the chirality of molecules at different extraterrestrial environments \citep{glavin2019}. 

So, how was the particular chirality of living organisms  chosen? Although many researchers argue that the choice is idiosyncratic and due to chance, there is a long history, going back to Abdus Salam \cite{1991JMolE..33..105S},  that associates it with the weak interaction – the one force of fundamental physics that expresses chirality. A quite different possibility is that chirality was imposed locally by a source of ultra-violet, circularly polarized light (UV-CPL), with sign which may or may not relate to the weak interaction. Laboratory experiments have shown asymmetric photolysis by UV-CPL in amino acids~\cite{deMarcellus}. However, the induced asymmetry is small (few percent), unlikely to lead to a homochiral state and some pre-biotic amplification mechanism is still required~\cite[{\it e.g.},][]{soai1995}. This suggests considering, instead, enantioselective bias in the evolution of living systems.  So, the other possibility is that  chirality is imposed at the biological level by the effects of cosmic radiation. {\bf Because they come from weak decays, muons are spin-polarized on average \cite{1993APh.....1..195L} and provide a direct link between the asymmetry of physics and the asymmetry of biology \cite{GB20}.}

 \section{Magnetically polarized muons as a chiral evolutionary pressure}

It has been proposed that spin-polarized cosmic radiation  can  induce asymmetric changes in helical biopolymers that may account for  the emergence of biological homochirality \cite{GB20}. Muons retain their polarization down to energies at which they can initiate enantioselective mutagenesis~\cite{GFB21}. Therefore, muons are  most likely to succeed in establishing the connection between broken symmetries in the standard model of particle physics  and that found in living organisms. 

 In \cite{GB20}, we showed that magnetically-polarized cosmic rays can impose a small, but persistent, chiral bias in the rate at which they induce structural changes in helical biopolymers, in particular, those that may have been the progenitors of RNA and DNA.  Enantioselective auto-catalysis, where molecules  of the same chirality  catalyze their own production while inhibiting the formation of their mirror-image \cite{frank1953}, has served as an important model because it exhibit some of the features of life, {\textit i.e.}, self-replication. Growth rates and mutation rates are correlated functions. When the growth rate is low, the probability to accumulate an adaptive mutation is strongly limited. 
Cosmic rays have a vital role, they promote mutations and therefore, they are  a driving force for evolution. We proposed that prebiotic chemistry produces both chiral versions of the molecular ingredients of life (\textit{i.e.} helical polymers capable of self-replication), and that at some stage in the earliest development of  biomolecules, a small difference in the mutation rate gives a chiral bias to one genetic polymer over its mirror-image \cite{GB20}.  {\bf A prediction of our model is that the mutation rate is dependent upon the spin-polarization of the radiation. }
 
  \begin{figure}
\centering
       \includegraphics[scale=0.25]{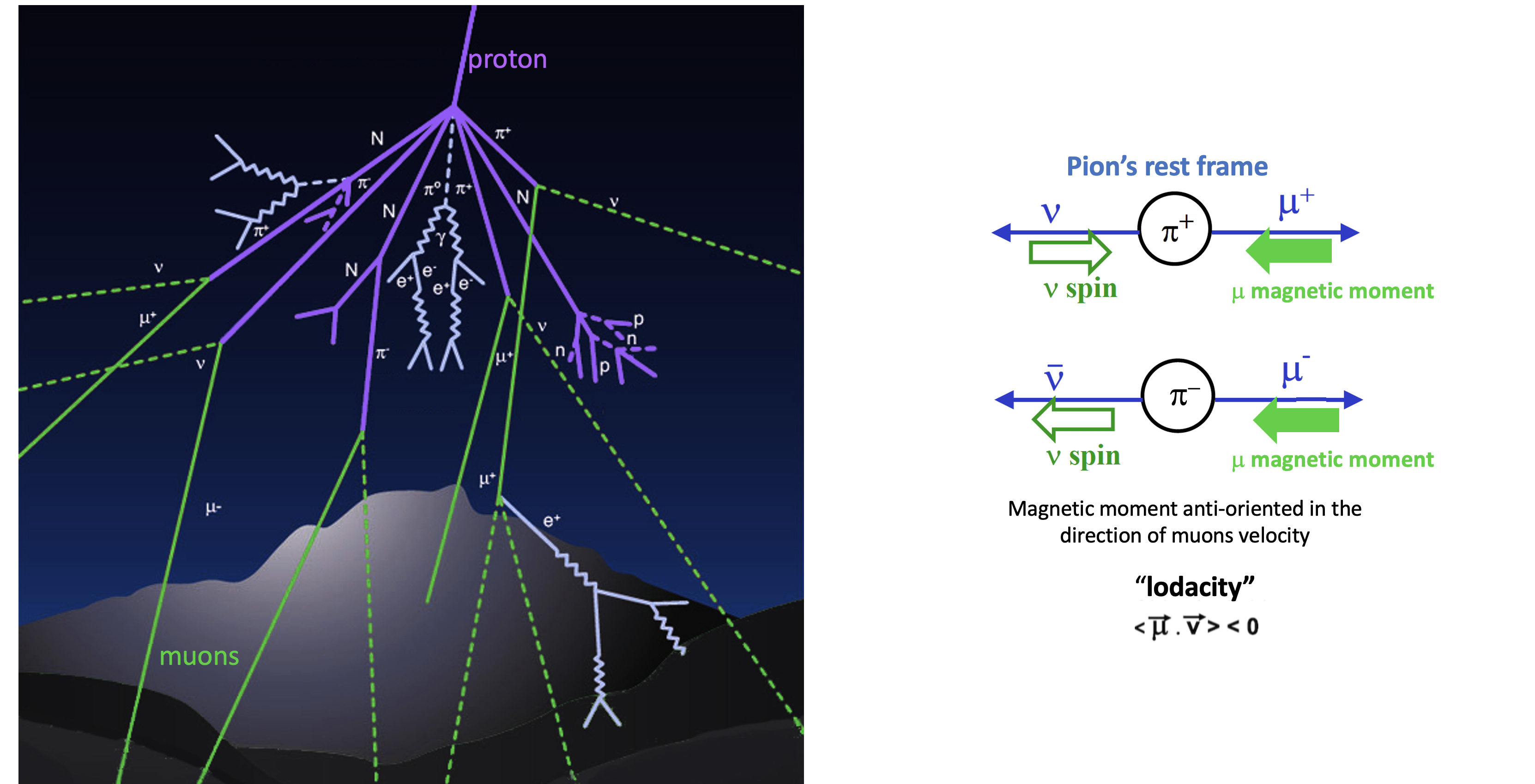}
\caption{The left panel of the Figure shows the development of a cosmic ray showers. We introduce a pseudoscalar quantity, ``lodacity'' (after lodestone) to express the physical chirality of the cosmic rays \cite{GB20}. This is defined by $
{\cal L}_i(T)=\overline{\hat{\boldsymbol{\muup}}\cdot\hat{\textbf v}}$,
where we average over all cosmic rays of type $i$ and kinetic energy $T$.  Well above threshold, ${\cal L}\sim-1$ for freshly created muons in the pion rest frame (right panel); it will be -0.3-0.6 in the observer frame, depending on the energy  \cite{1993APh.....1..195L, GFB21}.  ${\cal L}$ will be further degraded as the cosmic rays lose energy. In addition the secondary electrons will be rapidly depolarized as they lose energy and further diminish the lodacity of the cosmic rays that irradiate the molecules.}
\label{fig:muons}
\end{figure}

The chiral quantity associated with cosmic-ray showers is the lodacity ${\cal L}=\overline{\hat{\boldsymbol{\muup}}\cdot\hat{\textbf v}}$ \cite{GB20}.
Positive and negative muons  are produced in air showers mainly from charged pion decay. They are indicated in color in green in the sketch of Figure \ref{fig:muons}. The charge parity (CP) invariance leads to an universal sign of the cosmic-ray lodacity ${\cal L}<0$. 
At sea-level today, most cosmic rays are muons with an average flux $\sim160\,{\rm m}^{-2}\,{\rm s}^{-1}$  \cite{1993APh.....1..195L}. However, the flux and the atmosphere could have been quite different as we discussed in the first section and illustrated in the Figure~\ref{fig:doses}. The protobiological site, which we call the ``fount'', may have been below rock, water or ice which can change the shower properties and lodacity. Clay minerals, present in the fount, may have catalyzed the polymerization of the first biopolymers; they can also protect the  bases adsorbed on their surface from radiation \cite{biondi2007}. Deep-sea hydrothermal vents may not be the best candidate site for  polymerization of mononuleotides because they lack alternating dry and wet conditions which seem to be necessary for the emergence of biopolymers \cite{Damer20}. It has been argued that small, warm ponds, produced by hydrothermal conditions associated with volcanic activity on early Earth, are a good candidate environments for this process because their wet and dry cycles have been shown to promote the polymerization of nucleotides into long chains \cite{DaSilva15}. If life started shortly after the polymerization, then biological homochirality could also emerge in these small ponds, as they are more exposed to cosmic radiation than the very deep ocean.

\begin{figure}
\centering
       \includegraphics[scale=0.32]{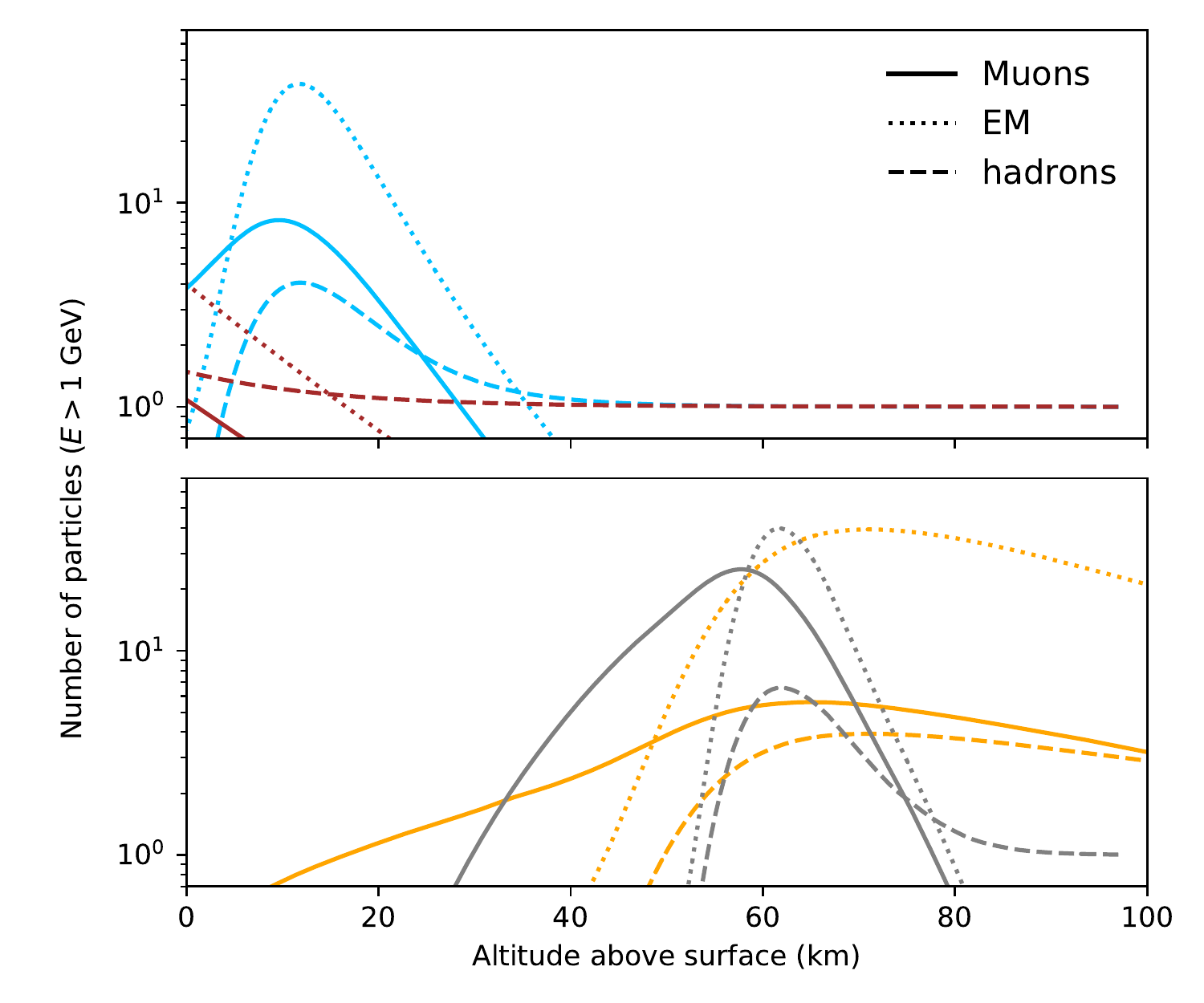}
       \includegraphics[scale=0.32]{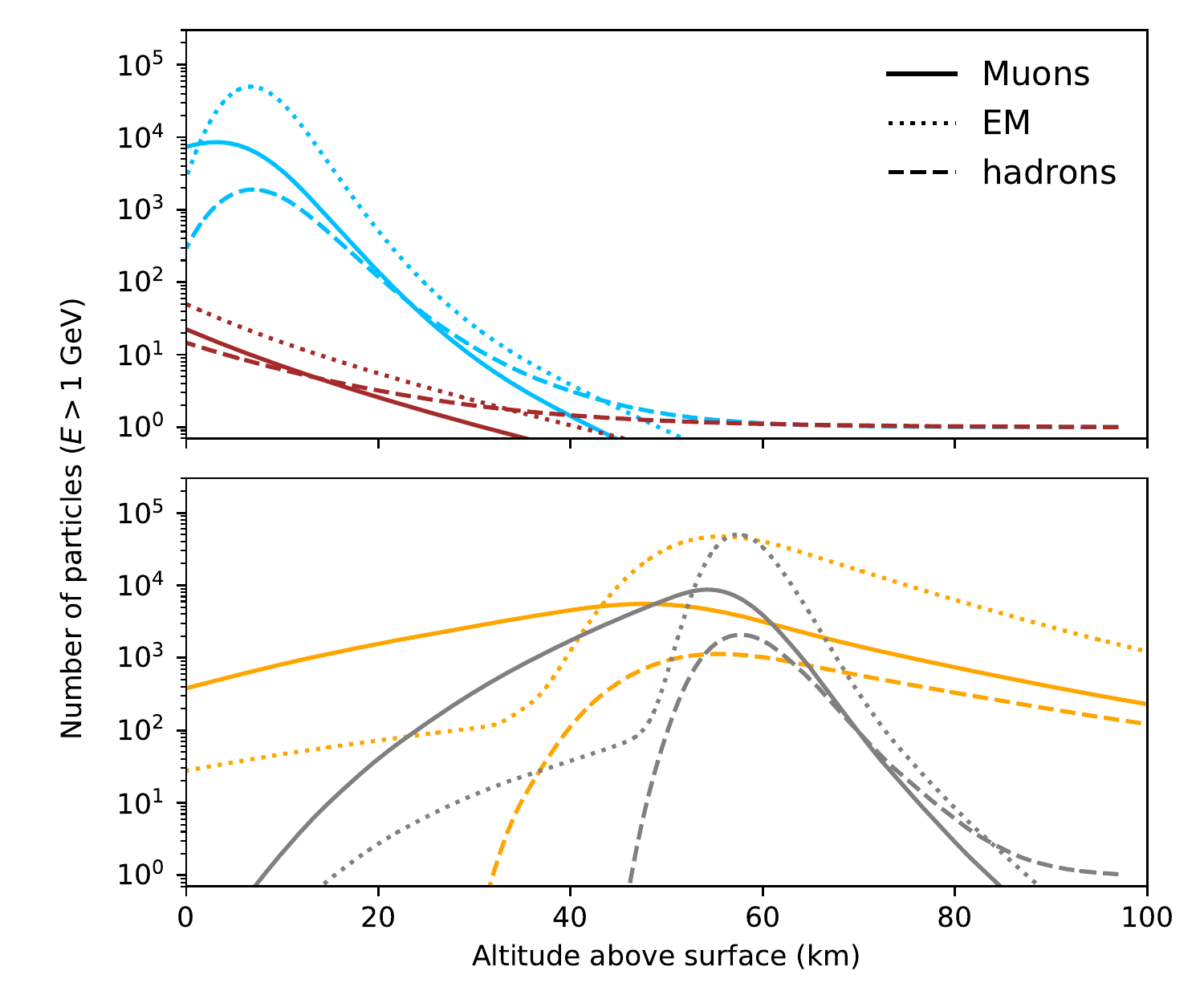}
    \includegraphics[scale=0.32]{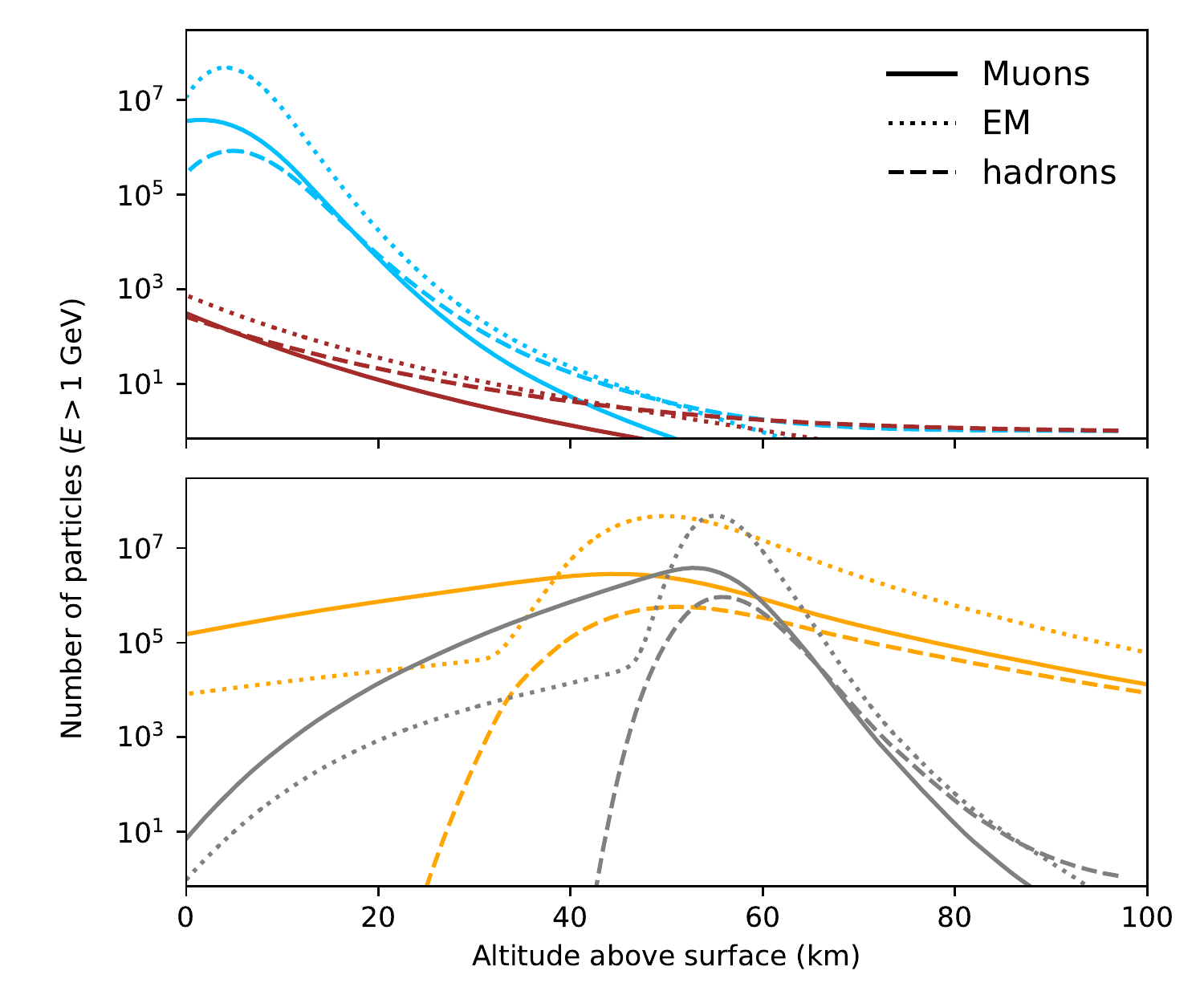}
\caption{Number of particles as a function of the altitude above the surface initiated by a proton at 1 TeV (left), 1 PeV (middle) and 1 EeV (right), on Earth (blue), Mars (red), Titan (yellow) and Venus (grey).}
\label{fig:founts}
\end{figure}

In \cite{GFB21}, we calculated the radiation doses deposited by muons at various prime targets for the searches of life in the solar system: Mars, Venus, Titan, icy moons and planetesimals.  Figure~\ref{fig:founts} shows the number of secondary particles on Earth, Mars, Titan and Venus, initiated by a primary cosmic ray proton at 1 TeV, 1 PeV and 1 EeV. One can see that the muons dominate at different altitude/depth, depending on the environment. On Mars, the atmosphere is thinner than on Earth, so the muons start to dominate only underground.   Titan and Venus are worlds with dense atmospheres.  On Venus, the habitable zone lies in the clouds, at $\sim50$ km above the ground where  muons are the dominant component of space radiation. Earth is unusual in that spin-polarized muons dominate the cosmic radiation at its surface. However, early Earth had an atmosphere with the same composition as modern Venus \cite{Sossi20}. {\bf The modeling of the polarized radiation in different (terrestrial and extra-terrestrial) environments seems to be a pre-requisite to understanding the conditions under which prebiotic molecules and the first living organisms can form. }

 \section{Proposed experiments}
 The idea that observed, biological homochirality is a consequence of the weak interaction expressed by cosmic rays (and ultimately of baryogenesis) is very attractive, but quite unproven. One way in which it could be falsified is if we discover extremophiles or ``alien'' life with the opposite chirality. Even if homochirality persists, we still do not have a mechanism that can confer a small chiral bias that is still large enough to be amplified by evolution. We have argued that any such bias would have to be carried by the muons not electrons. One way forward is to perform experiments on chiral molecules and living organisms. 
 
A possible experiment is to measure the mutation rate of two cultures of bacteria under spin-polarized radiation of different lodacity with energy above the threshold necessary to induce double strand breaks in DNA. {\bf If the coupling between lodacity and molecular chirality is efficient in introducing a chiral bias, one of the two cultures should exhibit a lower mutation rate~\cite{GB20}.} 
 
 We emphasize that much  can be learned experimentally from the comparison of chiral molecules involved in biology and using both signs of lodacity which can be created at accelerators. Once the dominant processes are identified, we can have confidence in our understanding of particle physics and quantum chemistry to draw the necessary conclusions. Additional experiments, relevant to our electromagnetic models, involve measuring the magnetic structure and properties of biopolymers. Experiments start to be performed  to see if it is possible to induce enantiomeric excesses in amino acids (using DL-alanine) with spin-polarized muons~\cite{TK19}. These experiments are ongoing. Experiments in the 70s showed inconclusive results~\cite{Lemmon74,Spencer79}. However, we now have the means to perform more sensitive investigations and perhaps be surprised by the results.

\bigskip
The research of NG was supported by the Simons Foundation. AF completed his
work as JSPS International Research Fellow (JSPS KAKENHI Grant Number 19F19750). We thank Stephen Blundell for useful discussions.

 \end{document}